\documentclass[runningheads]{llncs}

\usepackage{graphicx}
\usepackage{amsmath}
\usepackage{amsfonts}
\usepackage{amssymb}
\usepackage{mathtools}
\usepackage{multicol}
\usepackage{multirow}
\usepackage{xcolor}
\usepackage{subcaption}
\usepackage{placeins}
\usepackage{enumerate}

\usepackage{xcolor}
\newcommand{\reviewed}[1]{{\color[rgb]{0,0,0}{#1}}}

\newif\ifdraft
\draftfalse

\definecolor{orange}{rgb}{1,0.5,0}
\definecolor{pink}{rgb}{0.98, 0.38, 0.5}
\definecolor{darkgreen}{rgb}{0.055, 0.490, 0.016} 

\definecolor{senioryellow}{rgb}{0.86, 0.8, 0.19}
\definecolor{assistantblue}{rgb}{0, 0.133, 0.31}

\ifdraft
 \usepackage[normalem]{ulem}
 \newcommand{\RS}[1]{{\color{red}{\bf RS: #1}}}
 
 \newcommand{\PMN}[1]{{\color{orange}{\bf PMN: #1}}}

\else
 \newcommand{\sout}[1]{}
 \newcommand{\RS}[1]{{\color{red}{}}}
 
 \newcommand{\PMN}[1]{{\color{red}{}}}
 
\fi

\newcommand{\x}{\mathbf{x}}
\newcommand{\barx}{\bar{\x}}

\newcommand{\y}{\mathbf{y}}
\newcommand{\haty}{\hat{\y}}
\renewcommand{\d}{\mathbf{d}}
\newcommand{\D}{\mathcal{D}}

\newcommand{\comment}[1]{}

\newcommand{\ie}{{\it i.e.}}

\DeclareUnicodeCharacter{2212}{-}

\begin{document}
\title{CataNet: Predicting remaining cataract surgery duration}
%
%
\author{Andrés Marafioti\inst{1}, Michel Hayoz\inst{1}, Mathias Gallardo\inst{1}, Pablo Márquez Neila\inst{1}, Sebastian Wolf\inst{2}, Martin Zinkernagel\inst{2} and Raphael Sznitman\inst{1}}
\authorrunning{A. Marafioti et al.}

\institute{AIMI, ARTORG Center,  University of Bern, Switzerland \and
Department for Ophthalmology, Inselspital, University Hospital, University of Bern, Switzerland\\
\email{andres.marafioti@artorg.unibe.ch}}

\maketitle              
\begin{abstract}
Cataract surgery is a sight saving surgery that is performed over 10 million times each year around the world. With such a large demand, the ability to organize surgical wards and operating rooms efficiently is critical to delivery this therapy in routine clinical care. In this context, estimating the remaining surgical duration (RSD) during procedures is one way to help streamline patient throughput and workflows. To this end, we propose CataNet, a method for cataract surgeries that predicts \reviewed{in real time} the RSD jointly with two influential elements: the surgeon's experience, and the current phase of the surgery. We compare CataNet to state-of-the-art RSD estimation methods, showing that it outperforms them even when phase and experience are not considered. We investigate this improvement\reviewed{ and show that a significant contributor is the way we integrate the elapsed time into CataNet's feature extractor.} 
\end{abstract}


\section{Introduction}

Cataract surgery is one of the most common surgeries in the world, with over 10 million procedures conducted each year. Worldwide, 100 million people suffer from cataract-induced vision impairments and with the aging world population growing, the number of patients at risk of complete blindness is sharply increasing~\cite{Wang2017}. Yet, even though cataracts can easily be treated, the shear number of surgeries needed poses an organizational challenge of unprecedented scale.

At its core, cataract surgery involves using a surgical microscope to help replace a patient's eye lens, that has become opaque, with a synthetic clear lens. Depending on the risk of the patient~\cite{Achiron2016,lanza2020application} and the experience of the operating surgeon~\cite{devi2012,Schoeffmann2018}, the procedure can be performed in under 20~minutes, whereby the majority of delicate surgical phases last 6-15 minutes. In major outpatient cataract clinics, a single surgeon can operate over 50 patients in a given day. As such, the ability to streamline patients and prepare them for surgery plays an important role in surgical workflow and the organization around the operating room. In this context, the ability to appropriately estimate remaining surgical duration (RSD) is imperative to prepare the stream of upcoming patients and doing so as early as possible is critical.

To date, considerable efforts have been put into designing automated methods to predict RSD~\cite{Padoy2008,Franke2013,Guedon2016,Spangenberg2017,Maktabi2017,Bodenstedt2019,Rivoir2019}. Namely, \cite{Aksamentov2017}~presented the TimeLSTM network, which combined a CNN and an RNN to perform RSD prediction. This method, {which achieves}
good results for cholecystectomy surgeries, pre-trained its CNN for phase recognition thus requiring phase annotations. In an attempt to avoid this requirement, \cite{Twinanda2019}~introduced RSDNet which only used unlabeled surgical videos to predict the RSD. Relying on the implicit 
\emph{progress} label of the videos, the authors showed that either the surgical phase or progress labels could be effectively utilized for RSD~prediction on laparoscopic surgeries. In contrast to laparoscopic procedures however, no RSD methods have focused on cataract surgery. However, there is related research such as that of Neumuth et al.~\cite{Neumuth2012}, which proposed a surgical workflow management system potentially applicable to RSD estimation. Similarly, \cite{Zisimopoulos2018,Primus2018}~detected the current phase in cataract sequences from which RSD could be estimated. Yet these methods overlook important aspects: (1) surgeon experience plays a major factor in cataract surgery duration~\cite{devi2012,Schoeffmann2018} and (2) assessing the risk of the patient by inspecting the initial eye anatomy plays a key role in determining the difficulty and length of the procedure~\cite{Achiron2016,lanza2020application}.
\begin{figure}[t]
    \centering
    \includegraphics[width=.9\linewidth]{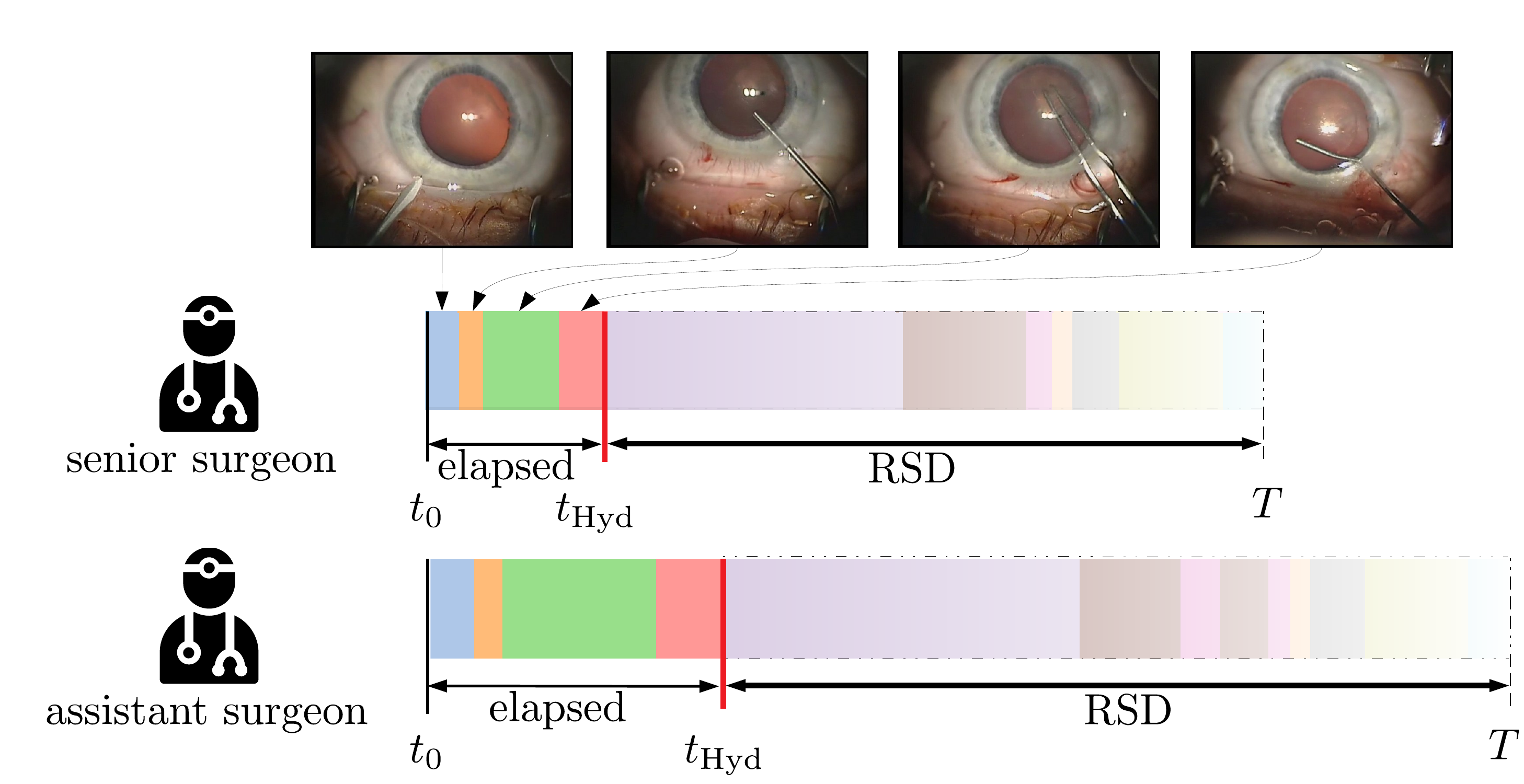}
    \caption{Surgeon experience and surgical phases play important roles in estimating remaining surgical duration (RSD) in cataract surgery. RSD predictions at $t_{\textrm{Hyd}}$ allows for optimal operating room patient management. }
    \label{fig:concept}
\end{figure}

In this work, we thus present a novel approach for online RSD prediction in cataract surgery. Our approach is to explicitly incorporate information from observed surgical phases, the operating surgeon's experience and the elapsed time at any given point to infer RSD prediction. We do this by embedding the video frames with the current elapsed time of the surgery, establishing a multi-task learning problem, and jointly identifying the surgeon's experience and the surgical phase, whereby overcoming a number of important limitations from recent methods (\ie,~RSDNet and TimeLSTM). By doing so, our approach avoids introducing additional complexities and yet considerably outperforms competing methods on both average RSD measures and RSD estimates at early stages of the surgery. In addition, we present an ablation study to identify the components of our method that give rise to the performance reported\footnote{Code and instructive examples are available at \url{github.com/aimi-lab/catanet}.}.

\section{Approach}

\subsection{Model}
Following~\cite{Schoeffmann2018}, we identify three key factors that influence the RSD: the surgeon's experience, the current surgical phase, and the elapsed time of the surgery (Fig.~\ref{fig:concept}). For accurate RSD estimation, it is thus critical that the predictive model is aware of these factors when processing the input video. To that end, we incorporate the factors into the model in a number of ways. The elapsed time, readily available at both training and inference time, is appended as an additional channel to the input video frames. On the other hand, surgeon's experience and surgical phase are unknown at inference time. Instead, we train the model to estimate them from the input data.
\begin{figure}[t]
    \centering
    \includegraphics[width=.8\linewidth]{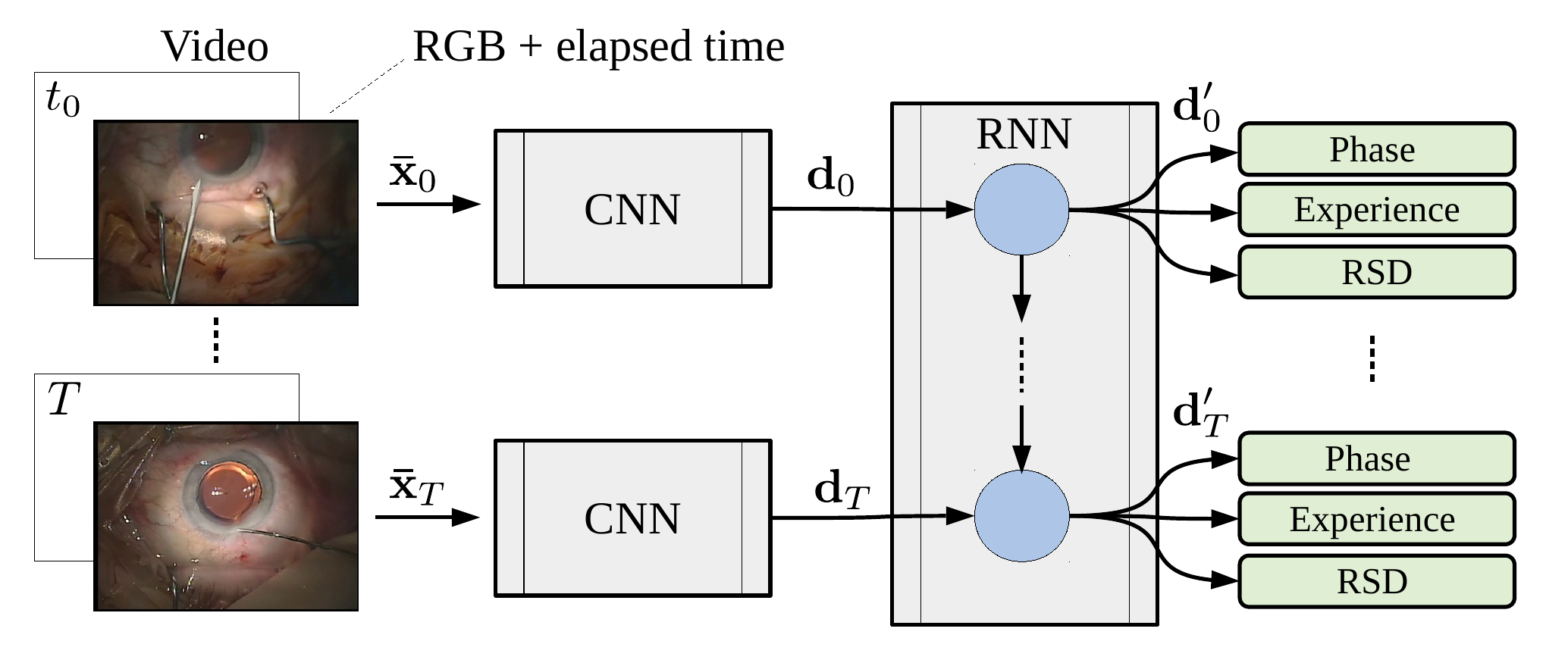}
    \caption{The end-to-end system. The inputs are video frames concatenated with the elapsed time of the surgery. Inputs are individually fed into the CNN and aggregated by the RNN, the output of which is finally passed through three independent fully connected layers to predict surgical phase, surgeon's experience, and RSD.}
    \label{fig:model}
\end{figure}

Fig.~\ref{fig:model}~depicts our model and how these three predictive factors are incorporated into it. Formally, our model consists of a CNN~$f:[0,1]^{3+1} \to \D$ that maps the input tensor~$\barx_t$ to a frame descriptor vector~$\d_t\in\D$, followed by a RNN~\cite{hochreiter1997long}$g:\D \to \D'$ that incorporates temporal information to produce a video descriptor vector~$\d'_t\in\D'$. We pass the input tensor~$\barx_t = \left[\x_t, \mathbf{1}\frac{t}{T_\textrm{max}}\right]$, which contains the input frame~$\x_t$ at time~$t$ and the elapsed time~$t$ as an additional channel, to the CNN. The elapsed time is scaled to the range~$[0, 1]$ by dividing~$t$ by the expected maximum video length~$T_\textrm{max}$ that we set to~$20\, \textrm{minutes}$. Passing the elapsed time at the image level enables the CNN to learn its embedding.

Every video descriptor vector~$\d'_t$ produced by the~LSTM is finally processed with three independent fully connected layers~($h^\textrm{exp}$, $h^\textrm{phase}$, $h^\textrm{rsd}$) to estimate the surgeon's experience~$\haty_t^\textrm{exp}$, 
the surgical phase~$\haty_t^\textrm{phase}$, 
and the RSD~$\hat{y}_t^\textrm{rsd}$. A softmax non-linearity is applied to obtain the probabilities $\haty_t^\textrm{exp}$ and~$\haty_t^\textrm{phase}$.

\subsection{Training objectives}

Our training dataset is a collection of tuples~$\left(\{\x_t\}_t, \{y^\textrm{rsd}_t\}_t, \{y^\textrm{phase}_t\}_t, y^\textrm{exp}\right)$ consisting of a video sequence~$\x_t$, the corresponding remaining surgical duration~$y^\textrm{rsd}_t$ per frame, surgical phases~$y^\textrm{phase}_t$ per frame, the surgeon's experience label~$y^\textrm{exp}$ per sequence. The index $t$~is the elapsed time of the sequence.

We use the labeled data to train our model by minimizing two different loss functions. First, the CNN~loss~$\ell_\textrm{cnn}$ is used to train the standalone CNN, without the~RNN, to classify the phase and experience of individual frames. To this end, we append two temporary linear layers, akin to $h^\textrm{phase}$ and~$h^\textrm{exp}$ above, acting on the output of the CNN~$\d_t$ to produce frame-level predictions~$\haty^\textrm{phase}_{\textrm{cnn},t}$ and~$\haty^\textrm{exp}_{\textrm{cnn},t}$. The CNN~loss minimizes the cross-entropies of both predictions,
\begin{equation}
\label{eq:l1}
    \ell_\textrm{cnn} = \textrm{H}(\haty^\textrm{phase}_{\textrm{cnn},t}, y^\textrm{phase}_t) + 
             \textrm{H}(\haty^\textrm{exp}_{\textrm{cnn},t}, y^\textrm{exp}_t).
\end{equation}
The RNN~loss~$\ell_\textrm{rnn}$, on the other hand, is used with video sequences to train the~RNN and to fine-tune the entire model end-to-end. It is a combination of the cross-entropies on phase and experience predictions, and the L1-norm of RSD~predictions,
\begin{equation}
\label{eq:l2}
    \ell_\textrm{rnn} = 
             \alpha\left|\hat{y}_t^\textrm{rsd} - y_t^\textrm{rsd}\right| +
             \textrm{H}(\haty^\textrm{phase}_t, y^\textrm{phase}_t) + 
             \textrm{H}(\haty^\textrm{exp}_t, y^\textrm{exp}_t),
\end{equation}
where the hyperparameter $\alpha$~weights the relative contribution of the L1-norm.

\section{Experiments}

\subsection{Training and test data}
We used the cataract-101 dataset~\cite{Schoeffmann2018} containing 101~videos (1'263'116~frames) with a resolution of $720\times{}540$~pixels acquired at 25~fps. \reviewed{We did not choose a minimum video length, but used every video in the dataset.} Each video is annotated with 10~surgical phases and the experience of the operating surgeon. Surgeries were performed by four different surgeons, divided in two senior surgeons (56~surgeries) and two assistant surgeons (45~surgeries). In addition, we manually labelled the start and end of each surgery, respectively, as the start of the first incision and the last tool interaction with the patient's eye.

The dataset was randomly split into 81~training and 20~test videos, so that 5~videos per surgeon remained in the test set. In the following experiments, we perform 6-fold cross-validation on the training split for model selection and hyper-parameter tuning. For inference, the output of all models is averaged.

\subsection{Implementation and baseline methods}
Our CNN uses a DenseNet-169~\cite{huang2017densely} architecture pre-trained on ImageNet. Input images are reshaped and cropped to $224\times224$, and the network produces descriptor vectors~$\d_t$ of 1664~dimensions. We implement our RNN as a LSTM~\cite{hochreiter1997long} with two layers of 128~cells, producing 128-dimensional video descriptor vectors~$\d'_t$.

Training is performed in four stages: (1)~\reviewed{First, to tackle class imbalance in surgical phases, we apply stratified sampling over the whole training dataset} and sample 8000~frames per phase. We train using the Adam optimizer with early stopping in all training stages. The CNN is trained to minimize~$\ell_\textrm{cnn}$ for 3~epochs with a learning rate of~$10^{-4}$, batch size of~100 and early stopping on sub-epoch validation loss. (2)~We minimize~$\ell_\textrm{rnn}$ to train the RNN on full video sequences, temporally downsampled to 2.5\,fps, for 50~epochs and a learning rate~$10^{-3}$. The weights of the CNN are frozen during this stage. (3)~The entire model is trained end-to-end minimizing~$\ell_\textrm{rnn}$. We apply truncated back-propagation on sub-sequences of 48~frames and setting the learning rate to~$5\cdot10^{-4}$ for 10~epochs. (4)~Finally, we fine-tune the RNN minimizing~$\ell_\textrm{rnn}$ for another 20~epochs while keeping the learning rate at~$5\cdot10^{-4}$. The weights of the~CNN are frozen during this stage. For the $\ell_2$~loss, we set~$\alpha=1$. We implemented our method with PyTorch~1.6 and trained models using two Nvidia GeForce GTX 1080 Ti~GPUs.

\reviewed{Given that no method for cataract RSD estimation exists, we compare our approach to two methods originally designed for laparoscopic surgery:}
\begin{description}
    \item[TimeLSTM~\cite{Aksamentov2017}:] A ResNet CNN trained for phase recognition, followed by a LSTM trained for RSD prediction.
    \item[RSDNet~\cite{Twinanda2019}:] A modified version of~\cite{Aksamentov2017}, where the CNN is trained for progress prediction and the elapsed time is concatenated to the LSTM's output.
\end{description}
Both methods were originally proposed for cholecystectomy surgeries and did not provide \comment{publicly available }implementations. Therefore, we use our own implementations for both baselines, following the respective publications.

We measure the quality of RSD predictions with the mean absolute error (MAE) per video,
$
    \textrm{MAE} = \frac{1}{T} \sum_{t=0}^{T-1} \left| \hat{y}^\textrm{rsd}_t - y^\textrm{rsd}_t\right|.
$
Similarly, we also provide MAE averaged over the last two (MAE-2) and five (MAE-5) minutes, as well as at the end of \textit{Hydrodissection} phase (MAE@Hyd). The latter metric is of clinical relevance in cataract surgery, as it highlights an appropriate time to prepare the following patient for surgery. In addition, we compute frame-wise accuracy (ACC) and F1-score per video to quantify surgical phase classification. 

\comment{We used three different reference methods: RSDNet~\cite{Twinanda2019}, TimeLSTM~\cite{Aksamentov2017}, and an oracle method which we devised. There is no public implementation of either RSDNet or TimeLSTM, so we implemented them following their respective publications to the best of our abilities and we provide these implementations in the supplementary material. Some details of the RSDNet which were unclear to us from the paper were clarified with the authors. Both networks were trained following the training schemes proposed in their respective papers, using the same 6-fold cross validation as for our method and averaging the solutions of every fold's model at testing time.

The last reference method is an oracle method which uses the ground truth of both the phase and experience labels. At the end of every phase, and dividing per surgeon's experience, we obtain the average RSD over the training set. Then, during testing, the oracle method predicts the RSD using the averages for the current surgical phase and experience.}


\subsection{Results}
\begin{table}[t]
    \caption{RSD prediction results. The MAE (mean$\pm$std in minutes) is shown for entire videos, the last two and five minutes, and at the end of \textit{Hydrodissection}.}
    \centering
    \setlength{\tabcolsep}{8pt} 
    \begin{tabular}{llccc}
    \hline
                 & Exp  &  CataNet  &  RSDNet & TimeLSTM \\
                 \hline
    \multirow{3}{*}{MAE@Hyd} 
    & All & $\mathbf{1.66 \pm 1.35}$ & $2.32 \pm 1.27$ & $2.34 \pm 1.54$\\
    \cline{2-5}
    & Senior & $\mathbf{1.22 \pm 0.97}$ & $2.86 \pm 1.31$ & $3.30 \pm 1.06$  \\
    & Assistant & $2.10 \pm 1.56$ & $1.78 \pm 1.02$ & $\mathbf{1.39 \pm 1.37}$ \\
    \hline
    \multirow{3}{*}{MAE-5} 
    & All & $\mathbf{0.64 \pm 0.56}$ & $1.37 \pm 0.83$ & $1.47 \pm 0.78$ \\
    \cline{2-5}
    & Senior & $\mathbf{0.78 \pm 0.60}$& $1.98 \pm 0.73$ & $2.06 \pm 0.70$  \\
    & Assistant & $\mathbf{0.51 \pm 0.23}$& $0.76 \pm 0.28$ & $0.88 \pm 0.14$ \\
    \hline
    \multirow{2}{*}{MAE-2} 
    & All & $\mathbf{0.35 \pm 0.20}$ & $1.23 \pm 0.53$ & $1.22 \pm 0.32$ \\
    \cline{2-5}
    & Senior  & $\mathbf{0.37 \pm 0.22}$ & $1.42 \pm 0.45$ & $1.43 \pm 0.32$  \\
    & Assistant & $\mathbf{0.34 \pm 0.18}$& $1.04 \pm 0.56$ & $1.03 \pm 0.13$  \\

    \hline \hline
    \multirow{3}{*}{MAE}   
    & All   & $\mathbf{0.99 \pm 0.65}$ & $1.59 \pm 0.69$ & $1.66\pm 0.79$  \\
    \cline{2-5}
    & Senior & $\mathbf{0.83 \pm 0.64}$& $1.97 \pm 0.73$ & $2.11\pm 0.70$  \\
    & Assistant & $\mathbf{1.15 \pm 0.65}$ & $1.19 \pm 0.36$ & $1.20\pm 0.59$ \\

    \hline
    \end{tabular}
    \label{tab:RSD-main}
\end{table}

\begin{figure}[t]

\begin{tabular}{ccc}
\includegraphics[width=0.4\linewidth]{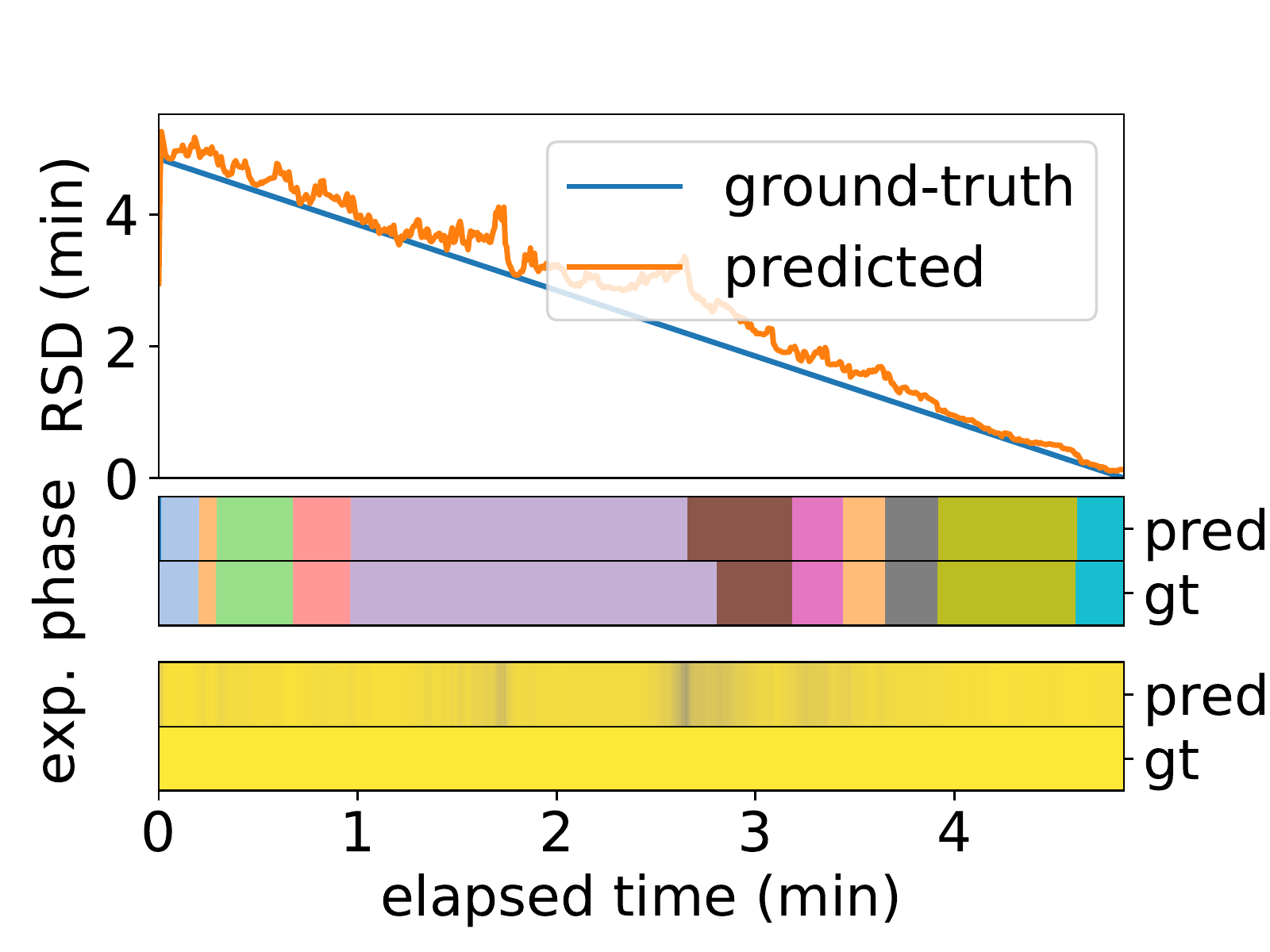} &
\includegraphics[width=0.4\linewidth]{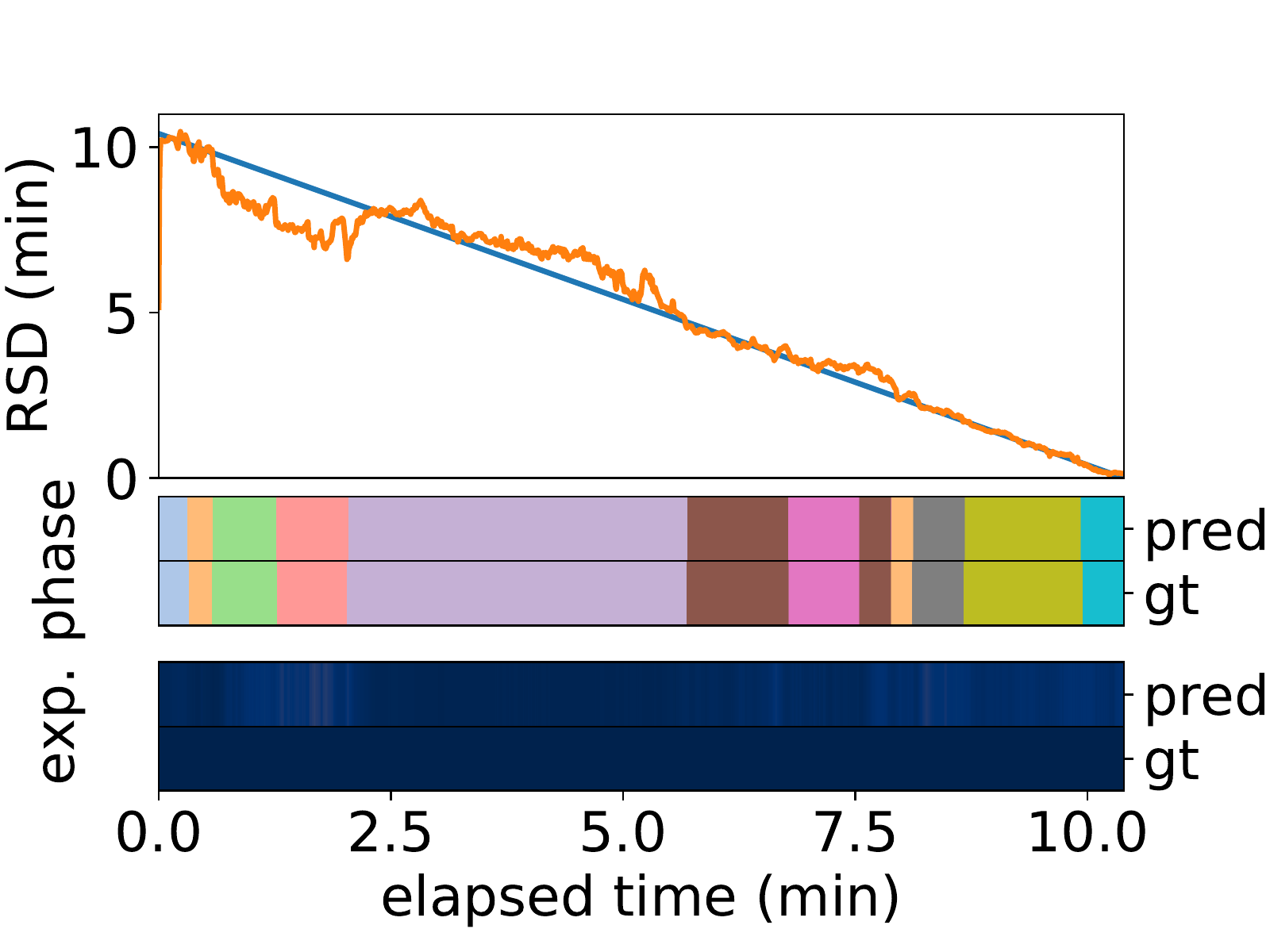} & 
\includegraphics[width=0.14\linewidth]{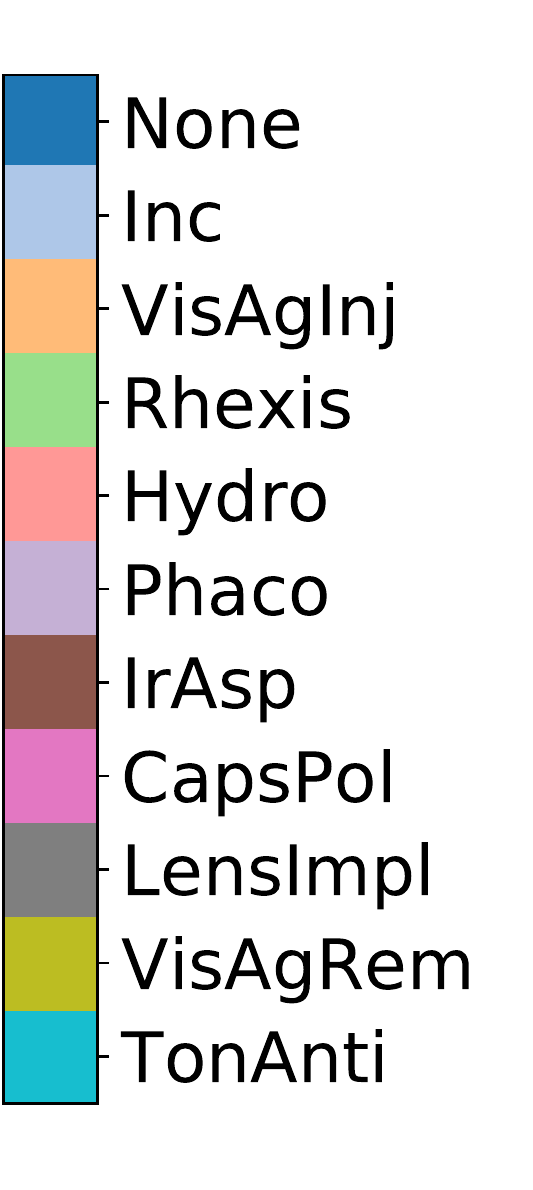} \\
\includegraphics[width=0.4\linewidth]{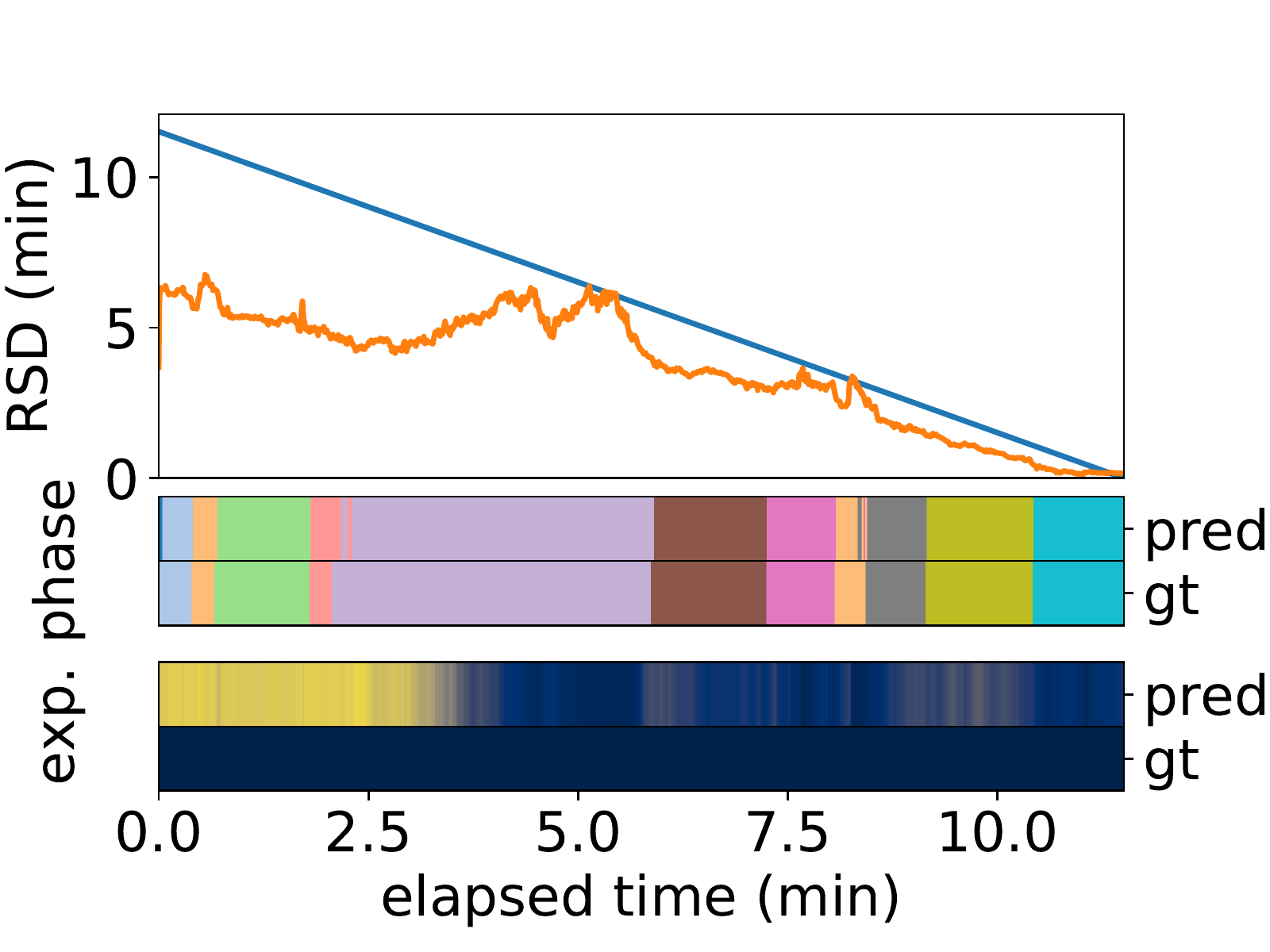} &
\includegraphics[width=0.4\linewidth]{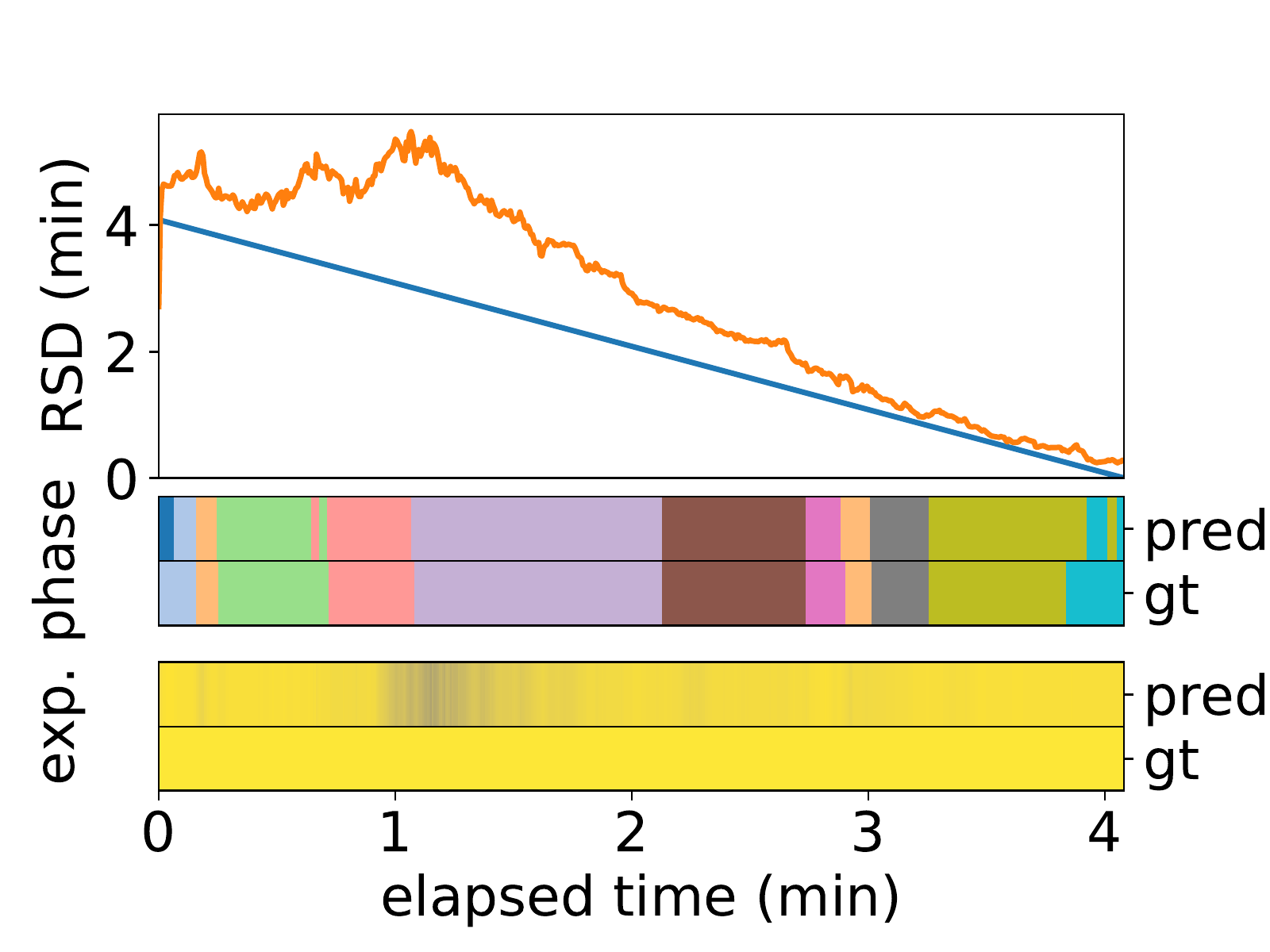} &
\includegraphics[width=0.115\linewidth]{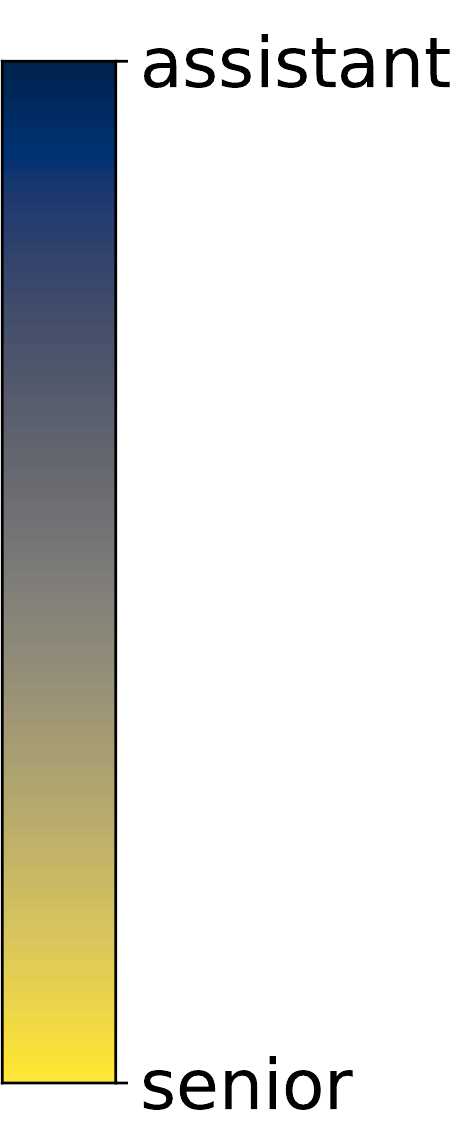}\\
\end{tabular}

\caption{Four examples of our method's outputs. For each plot, we show (\textbf{top})~the concordance between ground-truth and predicted RSD, (\textbf{middle})~the ground-truth and predicted surgical phases over time, and (\textbf{bottom})~the ground-truth and predicted probability of the surgeon's experience level.}
\label{fig:example_videos}
\end{figure}

Table~\ref{tab:RSD-main} shows the RSD prediction performance for all methods grouped by surgeon experience level. \reviewed{We group results by surgeon's experience level as both MAE and MAE@Hyd indirectly depend on the duration of the surgery and these take 5.6 and 11.8 minutes on average for senior and assistant surgeons, respectively.} CataNet outperforms both RSDNet and TimeLSTM in all but one metric. At the end of the critical \textit{Hydrodissection} phase, for all experiences CataNet performs 0.66~minutes better than RSDNet and 0.68~minutes better than TimeLSTM. At the end of this phase, CataNet is considerably better than the baselines for senior surgeons, but worse for assistant surgeons. Considering the prediction over the whole video, CataNet performs on average 0.6~minutes better than RSDNet and 0.67~minutes better than TimeLSTM. This can be explained by the fact that CataNet achieves comparable results for both senior and assistant surgeons. Overall, detection of the surgeon's experience is achieved with $0.92 \pm 0.16$ accuracy and can thus exploit the fact that senior surgeons show low variance in surgery duration\reviewed{, however we do not claim that this accuracy would translate to new surgeons}. The competing methods, on the other hand, tend to overestimate the duration of surgeries performed by senior surgeons.

We visualize CataNet's results for individual videos shown in Fig.~\ref{fig:example_videos} (see Supplementary material for more examples). Here, we see that predicting the surgeon experience \reviewed{on every frame} can be beneficial in determining the confidence in RSD predictions. That is, given that the experience of the surgeon is known by the operating staff, an incorrect classification in experience can serve as an easy and interpretable indicator when the system is performing poorly (\ie~overestimating the RSD for false \emph{assistant} predictions, or underestimating it for false \emph{senior} predictions). Additionally, considering that \emph{experience} is not a binary label, but a multi factored and scaled concept, our approach could be used to help assistant surgeons detect which phases of the surgery they could improve on. Finally, in two test set {sequences}, the surgeons fails to correctly perform the \textit{lens implantation} phase, leading to unexpected extensions of the surgeries by 2-3 minutes and consequently underestimate of RSD~before the mistake. However, our approach corrects the RSD~predictions shortly thereafter. Details of these two sequences can be found in the supplementary material.

{\bf{Ablation study: }}
CataNet is trained to classify the experience of the surgeon, the surgical phase, and the RSD, while its input is the video frames concatenated with the elapsed surgical time. To characterize the effects on performance of these different components, we show the performance of the following different approaches in Table~\ref{tab:ablation}: (i)~train the CNN to only predict surgical phases and the RNN to predict both phases and RSD; (ii)~train the CNN to only predict the surgeons experience and the RNN to predict both experience and RSD; (iii)~train the CNN and the RNN to estimate only the RSD; (iv)~same as (iii) but concatenate the elapsed time to the output of the LSTM (\ie,~as in RSDNet) instead of to the video frames. 
\begin{table}[t]
    \caption{Ablation evaluation for RSD prediction.}
    \centering
    \begin{tabular}{ll|c|ccccc}
    \hline
                & Exp &  CataNet  &  (i) phase & (ii) exp & (iii) RSD & (iv) elapsed \\
                 \hline
    \multirow{3}{*}{MAE@Hyd} 
    & All &  $1.66 \pm 1.35$ & $\mathbf{1.43 \pm 1.19}$ & $1.82 \pm 1.63$& $1.99 \pm 1.38$ & $2.28 \pm 1.34$ \\
    \cline{3-7}
    & Senior &  $\mathbf{1.22 \pm 0.97}$ & $1.42 \pm 1.38$ & $1.46 \pm 1.3$& $1.71 \pm 1.24$ & $1.45 \pm 0.59$ \\
    & Assistant &  $2.10 \pm 1.56$ & $\mathbf{1.43 \pm 1.05}$ & $2.18 \pm 1.91$& $2.26 \pm 1.52$ & $3.12 \pm 1.37$ \\
    \hline
    \multirow{3}{*}{MAE-5} 
    & All &  $\mathbf{0.64 \pm 0.46}$ & $0.74 \pm 0.56$ & $0.87 \pm 0.62$& $0.76 \pm 0.41$ & $0.75 \pm 0.34$  \\
    \cline{3-7}
    & Senior &  $\mathbf{0.78 \pm 0.60}$ & $0.88 \pm 0.73$ & $0.99 \pm 0.77$& $0.98 \pm 0.44$ & $0.85 \pm 0.31$ \\
    & Assistant &  $\mathbf{0.51 \pm 0.23}$ & $0.59 \pm 0.27$ & $0.76 \pm 0.43$& $0.55 \pm 0.24$ & $0.64 \pm 0.34$ \\
    \hline
    \multirow{3}{*}{MAE-2} 
    & All & $\mathbf{0.35 \pm 0.20}$ & $\mathbf{0.35 \pm 0.23}$ & $0.51 \pm 0.27$& $0.39 \pm 0.28$ & $0.44 \pm 0.20$ \\
    \cline{3-7}
    & Senior &  $0.37 \pm 0.22$ & $\mathbf{0.35 \pm 0.26}$ & $0.52 \pm 0.36$& $0.45 \pm 0.38$ & $0.51 \pm 0.22$ \\
    & Assistant &  $0.34 \pm 0.18$ & $0.36 \pm 0.21$ & $0.50 \pm 0.18$& $\mathbf{0.33 \pm 0.10}$ & $0.36 \pm 0.15$ \\
    \hline
    \hline
    \multirow{3}{*}{MAE} 
    & All &  $0.99 \pm 0.65$  & $\mathbf{0.98 \pm 0.58}$  & $1.22 \pm 0.92$ & $1.11 \pm 0.62$ & $1.34 \pm 0.73$  \\
    \cline{3-7}
    & Senior &  $\mathbf{0.83 \pm 0.64}$ & $0.91 \pm 0.77$ & $1.03 \pm 0.80$& $1.03 \pm 0.46$ & $\mathbf{0.83 \pm 0.30}$ \\
    & Assistant &  $1.15 \pm 0.65$ & $\mathbf{1.04 \pm 0.31}$ & $1.41 \pm 1.03$& $1.20 \pm 0.76$ & $1.85 \pm 0.67$ \\
\hline
    \end{tabular}
    \label{tab:ablation}
\end{table}

From these experiments, we can see that (i) generally performs as well as CataNet, even outperforming it for some metrics. However, CataNet generally achieves a better performance for senior surgeons, who conduct the bulk of actual cataract surgeries~\cite{Campbell2019}. In addition, we notice that (iii) performs better than (iv), showing that using the elapsed time as an input for the model considerably outperforms having it after the LSTM layer. Last, even when training without any labels, our approach (iii) performs better than that of RSDNet.

{\bf{Results on surgical phase classification: }}
Table~\ref{tab:phaseclass} shows CataNet's performance for phase classification. Compared to the state-of-the-art by Qui et al.~\cite{Qi2019}, CataNet achieves an increase of 12\% in accuracy from 0.84 to~0.95. Furthermore, CataNet reliably detects the \textit{Hydrodissection} phase, which is critical in the clinical context. Indeed, knowing the RSD at the end of this phase will improve the OR management since it corresponds to the moment where the next patient could be prepared for surgery.

\reviewed{{\bf{Inference speed: }}
RSD estimation is intended to be performed on real-time. We measured the execution time using a GeForce MX250 and avoided any overhead produced by other components of the system. We first run 100 frames through the GPU after which we measured the inference time on the next 1000 frames. The average time per frame was $34.3\pm1.9$ ms, which corresponds to $29.09$ fps. Considering that we sample the videos at 2.5 fps, we conclude that CataNet can easily be applied at 10 times real-time speed.}

\section{Conclusion}
We have proposed a novel \reviewed{real-time} method for estimating RSD for cataract surgeries from video feeds. Our approach jointly predicts the RSD, the surgeon's experience and the surgical phase, as these three elements are interconnected. Even when training our method without any labels, it outperforms the previous state-of-the-art RSD estimation models. We investigated the sources of this improvement and attribute these to (1) concatenating the video frames with the elapsed time and (2) including the phase and experience labels. \reviewed{Predicting the experience on every frame additionally increases the clinical applicability of our method by identifying low method confidence by observing predicted and real experience levels. Moving forward, a major challenge is in establishing large datasets to evaluate generalization capabilities and major clinical impact~\cite{bar2020impact}, for which assuring data consistency will be critical~\cite{Ghamsarian2020Deblurring}. In the future, we plan to investigate this and} how pre-operative data can be used to further improve RSD predictions for cataract surgery.
\begin{table}[t]
    \caption{Macro F1-score and micro accuracy averaged over the 6-fold models.}
    \centering
    \setlength{\tabcolsep}{8pt} 
    \begin{tabular}{lcccc}
    \hline
                 &  CataNet & Qui et al.~\cite{Qi2019} & TimeLSTM-CNN~\cite{Aksamentov2017} \\
                 \hline
    F1   & $\mathbf{0.93 \pm 0.06}$ & - &  $0.80 \pm 0.07$ \\
    F1-Hyd   & $\mathbf{0.94 \pm 0.08}$ & - & $0.84 \pm 0.17$  \\
    ACC & $\mathbf{0.95 \pm 0.05}$ & $0.84 \pm 0.06$ & $0.84 \pm 0.07$ \\
    \hline
    \end{tabular}
    \label{tab:phaseclass}
\end{table}

{\bf Acknowlegements:} This work was partially supported by the Haag-Streit Foundation and the University of Bern.

\bibliographystyle{IEEEtran}
\bibliography{cataracts, SSL, remaining_surgical_time}

\end{document}